\documentstyle[prl,aps]{revtex}
\input{psfig}
\draft
\begin{document}
\twocolumn[\hsize\textwidth\columnwidth\hsize\csname@twocolumnfalse\endcsname
\title{Continuum field description of crack propagation }
\author{I.S. Aranson,  V.A.  Kalatsky, and 
 V.M. Vinokur }
\address{Argonne National Laboratory, 9700 South Cass Avenue, Argonne, IL 60439}
\date{\today}
\maketitle

\begin{abstract}
We develop continuum field 
model for crack propagation in brittle amorphous solids. 
The model is  represented by 
equations for elastic displacements combined with 
the order parameter equation which accounts for the dynamics of defects.
This model 
captures all important phenomenology of crack propagation:
crack initiation, propagation, dynamic fracture instability,  
sound emission,
crack branching and fragmentation. 
\end{abstract}

\narrowtext
\vskip1pc]

The  dynamics of
cracks is the  long-standing
challenge in solid state physics and  materials science
\cite{pt,lawn}.
The phenomenology of the crack propagation is well-established
by recent experimental studies
\cite{exp1,sf,exp2,exp3,exp3_1,exp4,exp5}: once a flux of energy to the crack
tip passes the critical value, the crack becomes unstable, 
it begins to branch and emits sound.  
Although
this rich phenomenology is consistent with the
continuum theory, it {\it fails to describe it}  
because the way
the macroscopic object breaks
depends crucially on the details of cohesion on the microscopic scale
\cite{fm}.

Significant progress in understanding of
fracture dynamics was made by large-scale
(about $10^7$ atoms) molecular dynamics (MD) simulations
\cite{sim1,sim2}.
Although limited to sub-micron samples, these simulations were able to
reproduce several key features of the crack propagation, in particular, the
initial acceleration
of cracks and the onset of dynamic instability. However, detailed
understanding
of the complex physics of the crack propagation still remains a
challenge
\cite{science}.

The uniform motion of the crack is
relatively well-understood in the framework of the continuum theory
\cite{freund}.
Most of the studies treat cracks as
a front or interface separating broken/unbroken materials
and propagating under the forces arising from elastic stresses
in the bulk of material  and additional
cohesive stresses near the crack tip \cite{bm,abba,cln,fr}.
Although these investigations revealed some features of the
oscillatory crack tip instability, they are based on 
built-in assumptions,
e.g., on specific dependence of the fracture toughness on velocity,
structure of the cohesive stress etc.
To date there is no continuum model capable to describe in the same
unified framework the whole phenomenology of the fractures,  
ranging from
crack initiation to oscillations and branching.

In this Letter we present a continuum field theory of the 
crack propagation. Our model is  the wave  
equations for the elastic  deformations 
combined with the  equation for the order parameter, which is
related to the concentration of  material defects. 
The model
captures all important phenomenology:
crack initiation by small perturbation,  quasi-stationary 
propagation,  instability of  fast cracks,
sound emission, branching and fragmentation. 

{\it Model}. Our model is a set of the elasto-dynamic  equations 
coupled to the equation for order parameter $\rho$. 
We define the order parameter as 
the relative concentration of point defects in 
amorphous material (e.g.,  micro-voids). 
Outside the crack (no  defects) $\rho=1$ 
and $\rho=0$ inside the crack (all the atomic bonds are broken). 
At the crack surface $\rho$ changes from $0$ to $1$ on the scale 
much larger 
than the inter-atomic distance,  
justifying  the continuum description of the crack \cite{annil}.

\begin{figure}[h]
\centerline{ \psfig{figure=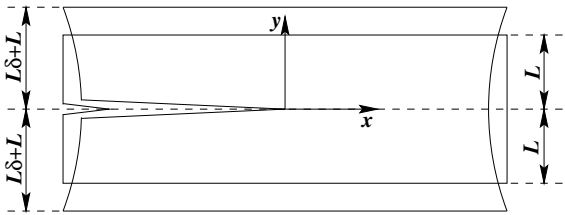,height=1.1in}}
\caption{Schematic representation of fixed-grips loading}
\label{Fig1}
\end{figure}

We consider the two-dimensional geometry 
focusing on the so-called type-I crack mode, see Fig. \ref{Fig1}. 
Equations of motion for an elastic media \cite{ll} are:
\begin{equation} 
\rho_0\ddot{u}_i=\eta \Delta \dot{u}_i+\frac{\partial 
\sigma_{ij} }{\partial x_j}
, \;\;j=1,2.
\label{elastic}
\end{equation} 
$u_i$ are the components of displacements, 
$\eta \Delta \dot{u}_i$ accounts for  viscous damping,  
 $\eta$ is the viscosity coefficient \cite{eta},  
$\rho_0$ is the density of material. 
In the following we set 
$\rho_0=1$.  
The stress tensor $\sigma_{ij} $ is related to deformations via
\begin{equation} 
\sigma_{ij}= \frac{E}{1+\sigma} \left( u_{ij} +\frac{ \sigma}{1- \sigma} 
u_{ll} \delta_{ij} \right) + \nu  \dot{\rho}  \delta_{ij}
\label{sigma} 
\end{equation}
where $u_{ij}$ is the elastic strain tensor,  $E$ is  the Young's modulus and 
$\sigma$ is the Poisson's ratio. To take into account effect of 
plastic deformations 
inside the material 
we introduce dependence  of $E$ upon $\rho$.  
For simplicity we consider   linear dependence 
$E=E_0 \rho$, where $E_0$ is the regular  Young's modulus. 
The  term $\sim \nu \dot{\rho}$ in 
Eq. (\ref{sigma}) accounts for the hydrostatic  pressure created 
due to generation of new defects, $\nu$ is the constant which can be obtained 
from comparison with the experimental data. Although this 
term can be formally 
interpreted  
as the hydrostatic 
pressure due to thermal expansion, 
we stress 
that the thermal expansion may  be not  the only mechanism 
contributing to the
magnitude of this term. 
There is experimental 
evidence that the temperature at the crack tip can be of the order 
of several hundred degrees \cite{temp}. However,   this would 
imply the concept of thermal equilibrium which is
unlikely to be  achieved at the crack tip. 


One can observe that Eqs.~(\ref{elastic}) 
are  linear  elasticity  equations  for $\rho=1$, i.e., outside 
the crack, and have trivial dynamics  
for $\rho=0$ (there is no dynamics inside crack). 

We assume that  
the order parameter $\rho$ is governed by  pure dissipative  dynamics 
which can be derived from the "free-energy" type functional ${\cal F}$, 
i.e., $\dot{\rho} = -\delta {\cal F}/\delta \rho$. 
Following Landau ideas on phase transitions \cite{ll1}, 
we adapt the simplest form for the "free energy" ${\cal  F} 
\sim \int dx dy ( D | \nabla \rho|^2 + \phi(
\rho))$, where the "local potential energy"  
$\phi $  has minima at $\rho=0$ and 
$\rho=1$. 
Choosing the polynomial form for $\phi(\rho)$ we  arrive at
\begin{equation} 
\dot{\rho} =D\Delta\rho-a\rho(1-\rho)F(\rho,u_{ll})+f(\rho)
\frac{\partial \rho}{ \partial x_l}\dot{ u}_l.
\label{op-eq}
\end{equation} 
Coupling to the displacement field  enters through the 
position of the unstable fixed point defined by the function 
$F(\rho,u_{ll})$, where $u_{ll}$ is the trace of the strain tensor. 
Constraint imposed on $F(\rho,u_{ll})$ is such that 
it must have  one zero in interval $1>\rho>0$: 
$F(\rho_c,u_{ll})=0$, $1>\rho_c>0$, and 
$\partial_\rho F(\rho=\rho_c,u_{ll})<0$. 
The simplest form of $F$ satisfying this constraint is 
$F(\rho,u_{ll})=1-(b-\mu u_{ll})\rho$, 
although any monotonic function of 
$\rho$ will show the qualitatively similar behavior. 
Here $\beta$ and $\mu$ are material constants 
related  to such  properties  as 
crack toughness and strain to failure. 
Coefficients 
$D$ and $a$ can be set to $1$ by rescaling 
$t\rightarrow a t$, $x_i\rightarrow \nu x_i$
where $\nu^2=D/a$.
 
The last term in Eq.~(\ref{op-eq})  represents the coupling of 
the order parameter to the velocity field $\dot{u}$   and  
is responsible for the localized 
shrinkage of the crack due to material motion. 
Since the specific form of the function $f(\rho)$ is  irrelevant, 
we take $f(\rho)=c \rho(1-\rho)$ 
to insure that 
$f$ vanishes at $\rho=0$ (no material)
and $\rho=1$ (no defects), 
where  $c$ is the dimensionless material constant. 
From our simulations we have found that this term in Eq. (\ref{op-eq}) 
is crucial to maintain  the sharp form of the crack tip. 

{\it Static solutions.} 
Eqs.~(\ref{elastic})-(\ref{op-eq}) have a dip-like solution 
corresponding to the open gap far behind the crack tip (a ``groove'' 
along $x$-axis for our geometry, see Fig.(1)). 
The static one-dimensional equations read:
\begin{equation}
\frac{\partial\rho u_{yy}}{\partial y}=0,\,\,\,
\frac{\partial^2\rho}{\partial y^2}-\rho(1-\rho)(1-(b-\mu
u_{yy})\rho)=0, 
\label{static-eqs}
\end{equation}
with the fixed-grips boundary conditions (BC): 
$u_y(y=\pm L)=\pm L\delta$ ($\delta$ --- relative displacement), 
$\rho(y=\pm L)=1$, and $\partial_y\rho(y=0)=0$. 
Exclusion of $u_{yy}$ from Eqs.~(\ref{static-eqs}) yields
\begin{equation}
u_{yy}=C/\rho,\;\;\partial_\xi^2 \rho=\rho(1-\rho)(1-\beta \rho), 
\label{rho1}
\end{equation}
where $C$ is a constant of integration 
($C\int^L_0dy/\rho(y)=L\delta$), $\beta=b/(1+\mu C)$, and 
$\xi = y\sqrt{1+\mu C}$. 
The solution to Eq.(\ref{static-eqs}) satisfying the BC 
for $L\to\infty$ is: 
\begin{equation}
\label{dip}
\rho=\frac
{\sqrt{(\beta+1)(1-\beta/2)}\cosh(\xi\sqrt{\beta-1})+2-\beta}
{\sqrt{(\beta+1)(1-\beta/2)}\cosh(\xi\sqrt{\beta-1})+2\beta-1},
\end{equation}
This solution exists for $1<\beta<2$. 
A deep and wide crack opening  is attainable 
if $2-\beta=\epsilon\ll 1$. 
In this case the BC for $u_y$ can be reduced to 
$\delta L = C(L+\pi\sqrt{3/(\epsilon b)})$ yielding an equation 
for $\epsilon$ since $C=(b/2-1)/\mu$. 
For  $\epsilon \ll 1 $ the width of the crack opening $d$, 
defined as $\rho(d/2)=1/2$, is $d=\sqrt{2/b}\ln(24/\epsilon)$. 
After exclusion of $\epsilon$ one arrives at
\begin{equation}
d = \sqrt{\frac{8}{b}}    \ln\left( \frac{\sqrt{8b}}{\pi}
L \left( \frac{2 \mu   \delta}{b-2} -1\right)  \right)
\label{d}
\end{equation}
The  solution to Eq. (\ref{d})
exists only if $\delta$  exceeds some critical value $\delta_c$
given by  $ \delta_c   \approx (b/2-1)/\mu $, which 
leads to the relation between strain to 
failure $\delta_c$ and the material parameters $\mu$ and $b$. 
The logarithmic, instead of linear, 
dependence of crack opening on system size  $L$ in Eq. (\ref{d}) is  a
shortcoming of the model  resulting  from oversimplified dependence 
of the function $F$ on $u_{ll}$. 


To study
the dynamics of cracks
we perform numerical 
simulations with Eqs. (\ref{elastic})-(\ref{op-eq}). 
We use explicit second-order numerical scheme with the  
number of grid points  up to $4000\times800$. 
Our model reproduces all important 
phenomenology: crack arrest below the critical stress, crack
propagation  above the critical stress, oscillations of the crack
velocity,  
crack branching and fragmentation. Selected results are presented 
in Figs. 2-5. 

\begin{figure}[h]
\centerline{ \psfig{figure=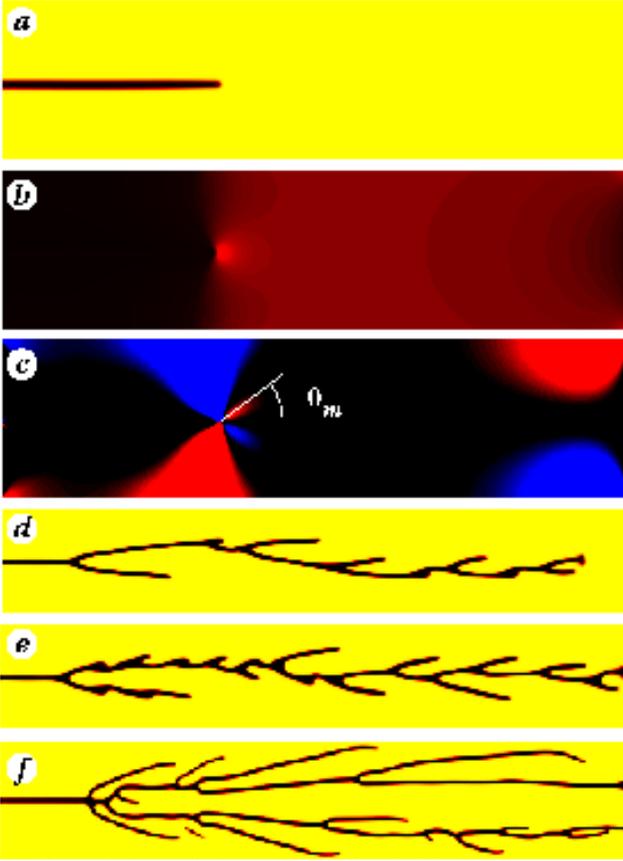,height=4.5in}}
\caption{Color-coded  images of  $\rho(x,y)$ (a);
hydrostatic pressure $p=-\rho (u_{xx}+u_{yy})$ (b); and 
shear $\rho (u_{xy}+u_{yx})$ (c).  
Domain size $800\times200$, number of grid points
$1600 \times 400$, and $\delta=0.05$, $\eta=0.25$, $E_0=10$, 
$\sigma=0.2, b=2.25$, 
$c=11$, $\nu=2.3$, $\mu=9.2$;
$\rho(x,y)$ for unstable propagation 
at  $\delta=0.089$ (d)  and $\delta=0.11$ (e). 
Domain size $1200\times200$, $2400\times 400$ grid points. 
(f) $\rho(x,y)$ for  propagation with fragmentation, 
$E_0=100, \sigma=0.36$, $c=16$, $\nu=2.3, \mu=54$ and $
\delta=0.03$, Domain size $2000\times400$, $4000\times 800$ grid points.
} 
\label{Fig2}
\end{figure}

\begin{figure}[h]
\centerline{ \psfig{figure=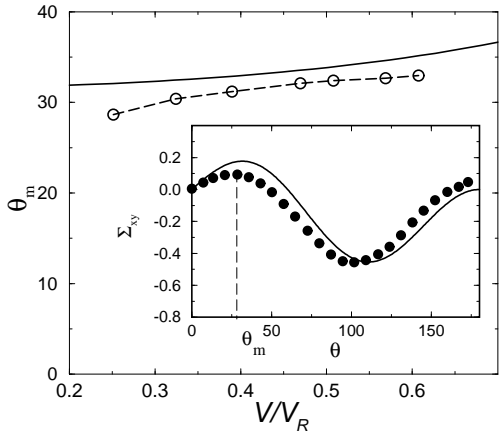,height=2.5in}}
\caption{$\theta_m$ vs  
crack velocity $V$ for  
parameters of  Fig. \protect\ref{Fig2}a-c. 
Solid line shows $\theta_m$ vs $V$ from 
linear elasticity \protect \cite{freund}, $\circ$ show result of 
numerical simulations. 
Inset: angular dependence of shear stress $\Sigma_{xy}$ (filled circles). 
The dependence from linear elasticity theory 
for infinite crack is shown in solid line. 
}
\label{Fig3}
\end{figure}

{\it Quasi-stationary propagation}. We considered crack propagation
initiated from a long notch with the length of the order $100$ units. 
At relatively small loadings $\delta$ we have observed a quasi-stationary 
propagation (no oscillations).  
The crack produces the stress concentration near the tip, 
and the stress  is relaxed behind  the tip, see
Fig. \ref{Fig2}b. The distribution of 
shear (Figs. \ref{Fig2}c,\ref{Fig3}) 
is close to that 
expected  from the elasticity theory. 
The angular dependence $\Sigma_{xy}(\theta)$
of the shear stress $\sigma_{xy}$ near the  tip  is close to the 
theoretical dependence $\sigma_{xy}(r,\theta) \sim r^{-1/2} \Sigma_{xy}^0
+...$,
where $\Sigma_{xy}^0(\theta) =
 \sin(\theta) \cos(3 \theta/2)$
obtained for the infinite stationary crack. The discrepancy 
can be attributed  to finite-size effects, velocity 
correction, etc. 
We computed the 
angle $\theta_m$ of the maximum shear stress vs 
crack speed $V$ normalized by the Rayleigh  speed $V_R$. 
(see Fig. \ref{Fig2}c, \ref{Fig3}).  
As one derives from the linear elasticity, the angle 
increases with the speed of the crack  \cite{freund}, in an agreement
with our numerical results.   

\vspace{-0.5cm} 

\begin{figure}[h]
\centerline{ \psfig{figure=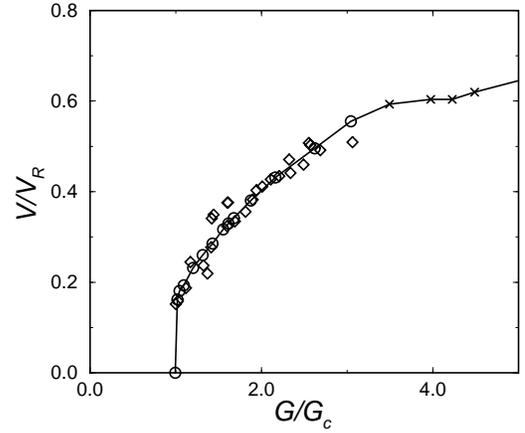,height=2.5in}}
\caption{Crack velocity 
$V/V_R$   vs
 $G/G_c$, where $G_c$ s the  onset value for  crack propagation.
$\circ$ correspond 
to stable propagation, $\times$ to unstable propagation, 
parameters the same as for Fig. \protect\ref{Fig2}a-e,
$\diamond$ show the experimental results from Ref. \protect 
\cite{exp2} normalized by  $V_R=1300$ m/s,  
the Rayleigh speed in PMMA in the  frequency range   of 0.6-1 MHz 
\protect \cite{exp3_1}. 
}
\label{Fig4}
\end{figure}

The calculated dependence of the crack  tip velocity  
on the  effective fracture energy
$G \sim\delta^2$, shown on Fig.~(\ref{Fig4}), 
demonstrates an excellent agreement  with the experimental data 
from Ref. \cite{exp2}. 
The instability of the crack occurs when the 
velocity becomes of the order of 55\% of the Rayleigh speed for 
the parameters of Fig.~\ref{Fig2}a-e. For 
parameters of Fig.~\ref{Fig2}f we have found lower value of 
the critical velocity, namely about 32\% of the Rayleigh speed. 
In all cases  the  instability 
manifests itself as pronounced velocity oscillations,  crack branching
and the sound emission from the crack tip. 

Our calculations indicate absence of the 
minimal crack velocity, the so-called velocity gap \cite{fm}. 
The initial velocity jump, seen experimentally as well as in some 
of our simulations (see Fig.~\ref{Fig4}), 
is attributed to the fact that the initial crack (notch) 
is too short or too blunt.

{\it Instability of cracks propagation}. 
Taking the sufficiently 
large  values of $\nu$ and $c$ and starting from short 
cracks with the large load  we observed 
consecutive crack branching. Since Eqs. (\ref{elastic})-(\ref{op-eq})
are homogeneous, these secondary crack branches typically retract 
after the stress at the tip of the shorter crack relaxes. Although
this retracting  may indeed 
take place, e.g., in vacuum the small cracks may 
heal, the oxidation of the crack surface and lattice trapping  
would prevent cracks from healing. In order to model these effects,  one 
can introduce an additional field representing concentration of oxygen 
and then couple it to the order parameter. 
In some simulations, we multiplied r.h.s. of Eq. (\ref{op-eq})
by a monotonic 
function $w(x-x_{tip})$: $w(x>0)=1 $  and $w(x\to-\infty)\to 0$, 
where $x_{tip}$ is the crack tip position. 
Thus, we slowed down the evolution of $\rho$
behind the crack tip, 
which, in turn, prevents secondary cracks from healing. 
We succeeded to obtain  realistic 
crack forms, see Fig. \ref{Fig2}d,e. 
For fast cracks  the ``freezing'' 
is not a necessity, since the 
retraction is rather slow.
Fig.~\ref{Fig2}f shows results without freezing: 
massive crack branching along with crack 
healing are present.

\begin{figure}[h]
\centerline{ \psfig{figure=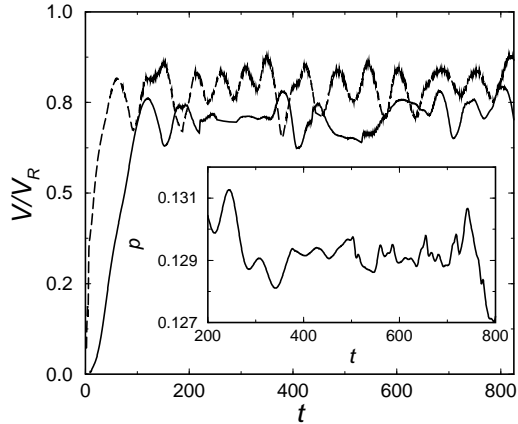,height=2.5in}}
\caption{The crack tip velocity $V$ 
normalized by Rayleigh velocity $V_R$ vs time 
for $\delta=0.089$ (solid line) and $\delta=0.11$ (dashed line), 
parameters of Fig. \protect \ref{Fig2}d,e.
Inset: oscillations of pressure $p$ far away from the crack tip
for $\delta=0.089$.}
\label{Fig5}
\end{figure}

Far away from 
the crack tip we have registered oscillations of hydrostatic pressure 
(see Fig. \ref{Fig5}, Inset), which is a clear indication of the sound 
emission by the crack tip. The sound waves 
reflected from boundaries  
may also induce  velocity oscillations, but they 
do not provide a mechanism for branching \cite{foot}. 
Increase in the applied displacement $\delta$ 
results in increase 
of amplitude 
and the number of secondary branches  
(cm. Figs. \ref{Fig2}d-f and \ref{Fig5}).

Some estimates are in order. Our unit of length $\lambda$ is the 
width of the craze zone and is of the order of a micron in PMMA\cite{annil}. 
The unit of time $\tau$ is obtained from $\lambda\sim1\mu$ and Rayleigh 
speed $V_R\sim10^3\mbox{m/s}$, which gives $\tau \sim \sqrt{E_0} 10^{-9}$
$s$.
In experiments \cite{sf,exp2,exp3,exp3_1,exp4} the characteristic time of 
velocity oscillations $\tau_0$ is of the order of 1 $\mu s$. 
Our model  gives 
$\tau_0 \sim 10^2 \tau \sim 0.1$ -- $1 $ $\mu s$ for $E_0=10$ -- $100$.

{\it Conclusion}. We have developed a continuum field theory of the
crack propagation. The central element of our approach is 
the  description of  crack by the order parameter. 
The proposed approach enables us to avoid 
the stress singularity at the crack tip
and to derive the  tip instability. 
Our model is complimentary to 
MD simulations of cracks and 
allows for a  description of fracture  phenomena on large scales.  
The parameters of our model 
can be obtained from comparison
with the experiment. 
It will be interesting  to derive 
the  order parameter equation from 
discrete models of crack propagation \cite{marder1,kessler}. 

We are grateful to M. Marder, H. Swinney, 
J. Fineberg, V. Steinberg, 
H. Levine, E. Bouchaud, A. Bishop, I. Daruka for 
stimulating discussions.
This research is supported by
US DOE, grant W-31-109-ENG-38.


\references
\bibitem{pt} M. Marder and J.   Fineberg,  Physics Today {\bf 49}, 24 (1996)
\bibitem{lawn} B. Lawn, {\it Fracture in Brittle Solids} (Cambridge
University Press, Cambridge, 1993), 2nd ed. 
\bibitem{exp1} J. A. Hauch, D. Holland, M. P. Marder, and H. L. Swinney, 
\prl { \bf 82}, 3823 (1999)
\bibitem{sf} E. Sharon and J. Fineberg, Nature {\bf 397}, 333 (1999) 
\bibitem{exp2} E. Sharon, S. P. Gross, and J. Fineberg, Phys. Rev. Lett. 74, 5096 (1995); {\it ibid} {\bf 76}, 2117 (1996). 
\bibitem{exp3} J.F. Boudet, S. Ciliberto, and V. Steinberg, 
Europhys. Lett. {\bf 30}, 337 (1995)
\bibitem{exp3_1} J.F. Boudet, S. Ciliberto, and V. Steinberg, 
 J. Phys. II (France) {\bf 6}, 1493 (1996)
\bibitem{exp4} J. Fineberg, S. Gross, M. P. Marder, and H. L. Swinney
\prl {\bf  67}, 457 (1991) 
\bibitem{exp5} D. Daguier, S. Henaux, E. Bouchaud, and F. Creuzet \pre {\bf 53},
5637 (1996)
\bibitem{fm} 
J. Fineberg and M. Marder,
Phys. Reports {\bf 313}, 1 (1999)
\bibitem{sim1} 
F. Abraham, \prl {\bf 77}, 869 (1997); 
 S. J. Zhou, D. M. Beazley, P. S. Lomdahl, 
and B. L. Holian, \prl {\bf 78},479 (1997); R. K. Kalia, A. Nakano, K. Tsuruta, and P. Vashishta, \prl  {\bf 78} 689 (1997); 
D. Holland and M. Marder \prl {\bf 80}, 746 (1998); 
F. Cleri, S. Yip, D. Wolf, S. R. Phillpot, \prl {\bf  79}, 1309 (1997)
\bibitem{sim2}Computing in Science \& Engineering,
{\bf 1} (1999) dedicated to dynamic fracture analysis. 
\bibitem{science} {Cracks: More Than Just a Clean Break}, Special Section in 
Science {\bf 281}, 943 (1998). 
\bibitem{freund} L.B. Freund, {\it Dynamic Fracture Mechanics} (Cambridge
University Press, N.Y., 1990). 
\bibitem{bm} E.A. Brener and V.I. Marchenko, \prl {\bf 81}, 5141 (1998)
\bibitem{abba} 
M. Adda-Bedia and M. Ben Amar, \prl {\bf  76}, 1497 (1996);
M. Adda-Bedia, R. Arias,  M. Ben Amar, and F. Lund, 
\prl {\bf 82}, 2314 (1999)
\bibitem{cln} 
E. S. C. Ching, J. S. Langer, and H. Nakanishi, \pre {\bf  53}, 2864 (1996).
\bibitem{fr} S. Ramanathan and D. S. Fisher, \prl  {\bf 79}  877 (1997)
\bibitem{annil} The characteristic scale of $\rho$ is related to the 
damage zone width   on the crack surface and is 
of the order of  a micron
for brittle polymers (the craze layer width, 
R.P. Kambour, J. Polymer Sci. A-2, {\bf 4}, 17 (1966)).   
\bibitem{ll} L.D. Landau and E.M. Lifshitz, {\it Theory of Elasticity}, 
(Pergamon Press, Oxford, 1964). 
\bibitem{eta} The specific  form of the 
dissipative terms does not change qualitatively the mechanism of 
crack tip instability. 
\bibitem{temp} K. N. G. Fuller, P.G. Fox and J.E. Field,
Proc. R. Soc. Lond. A {\bf 341}, 537 (1975);
J. A. Kallivayalil and A. T. Zender, Int. Jour. of Fracture
{\bf 66}, 99 (1994)
\bibitem{ll1} L.D. Landau and E.M. Lifshitz, {\it Statistical Physics}, 
(Pergamon Press, Oxford, 1980). 
\bibitem{foot} 
The interaction with 
sound can be reduced by introducing additional  
 damping term $\sim \dot{\bf u}$ to the rhs of Eq. 
\protect (\ref{elastic}).
\bibitem{marder1} M. Marder, \pre {\bf 54} 3442 (1996)
\bibitem{kessler} 
D.A. Kessler and H.  Levine,  \pre {\bf 59},  5154 (1999) 
\end{document}